\documentclass[10pt,final,twocolumn,twoside]{IEEEtran}
\usepackage{framed}
\usepackage{graphicx} \usepackage{amsmath} \usepackage{amssymb}
\usepackage{color}
\usepackage{amssymb}
\usepackage{amsthm} 
\usepackage[linesnumbered,ruled,vlined]{algorithm2e}
\SetKwComment{Comment}{$\triangleright$\ }{}
\newlength\algowd

\usepackage{cite} 
\usepackage[latin1]{inputenc}
\usepackage[caption=false,font=footnotesize]{subfig} 
\usepackage{stfloats}
\usepackage{url}
\usepackage{mathtools}
\usepackage{varwidth}

\usepackage{xpatch}
\makeatletter
\patchcmd\@makecaption{\\}{.~}{}{\fail}
\makeatletter
\DeclareMathOperator{\getNbr}{getNeighborAccessBeamPower}
\DeclareMathOperator{\getSer}{getServingAccessBeamPower}
\DeclareMathOperator{\getNodeId}{getNodeId}
\DeclareMathOperator{\getServingNode}{getCurrentServingNode}

\DeclareMathOperator{\getTTTValue}{getTimeToTriggerValue}
\DeclareMathOperator{\startTTTtimer}{startTimeToTriggerTimer}

      \DeclareMathOperator*{\argmax}{arg\,max}    

\newcommand{\mb}[1]{\mbox{#1}}
\newcommand{\norm}[1]{\left\lVert#1\right\rVert}
\newcommand{\mbf}[1]{\mathbf{#1}}     \newcommand{\mbs}[1]{\boldsymbol{#1}}
\newcommand{\RPT} {\mbs{\mathcal{R}}}
\newcommand{\ACCNBR} {\mbf{b}_{\mbox{\scriptsize nbrs}}}
\newcommand{\ACCSER} {\mb{b}_{\mbox{\scriptsize serv}}}
\newcommand{\BMAX} {\mb{b}_{\mb{\scriptsize max}}}

\newcommand{\TTTVAL} {\beta_{v}}

\begin{document}

\title{5G Handover using Reinforcement Learning}
\date{\today}
\author{Vijaya~Yajnanarayana, Henrik Ryd\'{e}n, L\'{a}szl\'{o} H\'{e}vizi\\
  Ericsson Research\\
  Email: vijaya.yajnanarayana@ericsson.com}
\maketitle
\thispagestyle{empty}
\pagestyle{empty}

\begin{abstract}
In typical wireless cellular systems, the handover mechanism involves reassigning an ongoing session handled by one cell into another.  In order to support increased capacity requirement and to enable newer use cases, the next generation wireless systems will have a very dense deployment with advanced beam-forming capability. In such systems, providing a better mobility along with enhanced throughput performance requires an improved handover strategy. In this paper, we will detail a novel method for handover optimization in a 5G cellular network using reinforcement learning (RL). In contrast to the conventional method, we propose to control the handovers between base-stations (BSs) using a centralized RL agent. This agent handles the radio measurement reports from the UEs and choose appropriate handover actions in accordance with the RL framework to maximize a long-term utility. We show that the handover mechanism can be posed as a contextual multi-armed bandit problem and solve it using Q-learning method. We analyze the performance of the methods using different propagation and deployment environment and compare the results with the state-of-the-art algorithms. Results indicate a link-beam performance gain of about $0.3$ to $0.7~\mb{dB}$ for  practical propagation environments. \\

\end{abstract}
\section{Introduction}
\label{sec:intro}

\begin{figure}[t]
  \centering
  \fbox{\includegraphics[width=3.2 in]{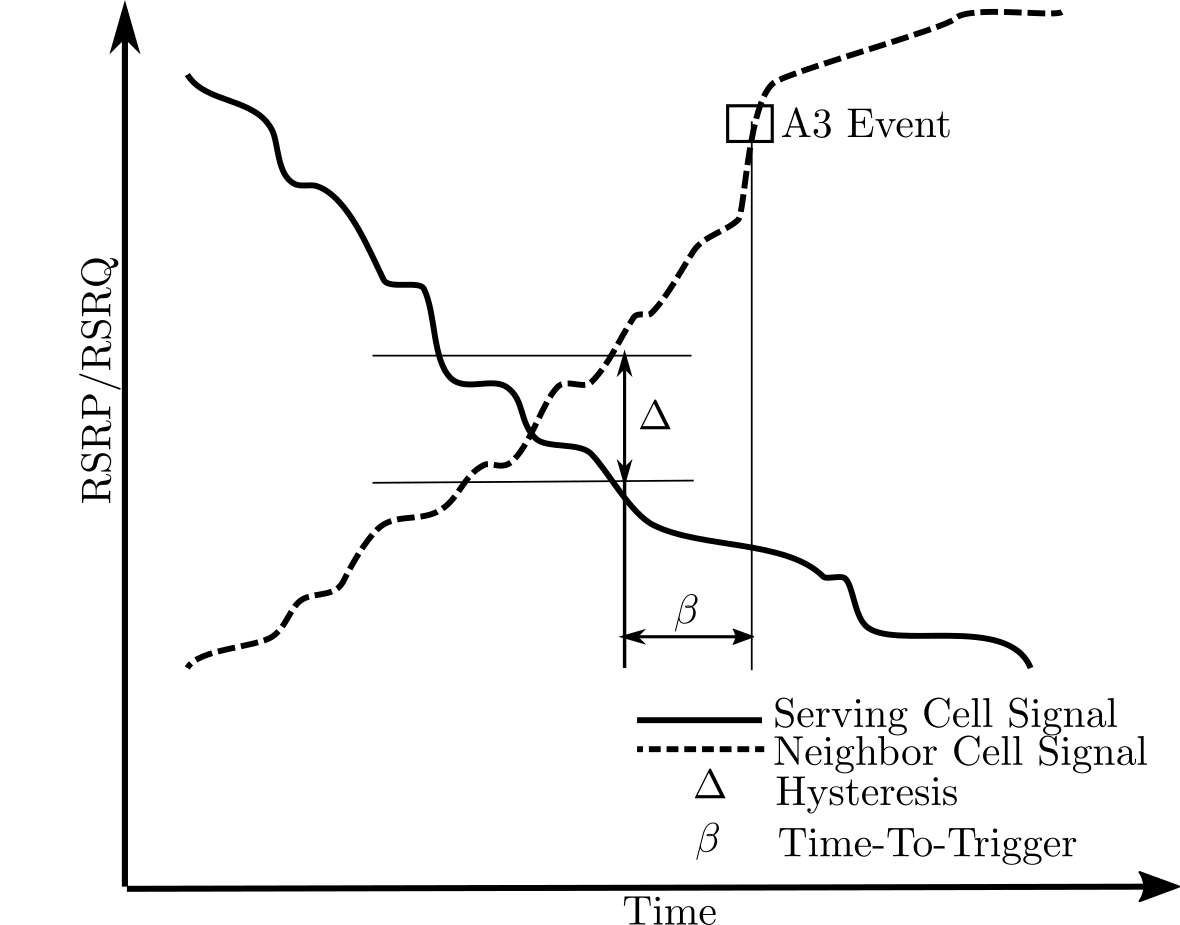}}
  \caption{Illustration of the HO mechanism in cellular networks}
  \label{fig:handover}
\end{figure}

In cellular wireless systems, mobility is achieved through handover (HO) mechanism. This enables UEs to move seamlessly within the coverage area of the network. The HO mechanism involves reassigning an ongoing session handled by one cell into another. A UE in the network will either be in an idle or connected mode. In idle mode, the UE just camps into a cell and does not have any active signaling or data-bearers to the base-stations (BSs) . However, when in connected mode, the BS will allocate resources to the UEs and there will be an active signaling on the data and control channels. In this paper, we describe a novel technique for connected-state intra-frequency HOs in 5G context. In typical cellular networks, UEs continuously monitor the signal strengths of the serving and neighbor cells, and report them to the serving base station. To illustrate this, consider a UE moving away from the serving cell near the cell edge. As shown in Fig.~\ref{fig:handover},  when the serving cell reference signal received power (RSRP) decreases below the acceptable level, and the neighbor cell RSRP is higher than the serving cell by a threshold (hysteresis value), then the serving BS initiates a HO. The RSRP measurements are typically done on the downlink reference signals. This algorithm is discussed in more detail in  \cite{TS36331,handover-book}. The hysteresis value ($\Delta$) along with time-to-trigger\footnote{The duration for which the target cell RSRP is above serving cell by $\Delta$ (Refer to Fig~\ref{fig:handover})} ($\Gamma$) is used to overcome the ping-pong effect.

\subsection{Related Work}
\label{ss:rw}
There exist several algorithms which computes the HO parameters  such as time-to-trigger ($\Gamma$) and hysteresis value ($\Delta$) optimally.  The algorithms in \cite{Leu-1} and \cite{Leu-2} discuss methods to overcome the ping-pong effect during HO. Optimization of HO between macro and femto BS by exploiting the UE information such as velocity, RSSI, etc. is discussed in \cite{Wu}.

Machine Learning has been proposed in several HO optimization problems. In hybrid cellular networks consisting of drone and terrestrial UEs, the main-lobe of a BS antenna is down-titled to serve terrestrial UEs. This results in drone UEs being frequently served by the side-lobes of the BS antennas \cite{Drone1}.  This creates a fragmented coverage area served by different BSs, thus increasing the radio link failures (RLFs)  and ping-pongs \cite{Drone2}. In \cite{DroneMobility}, the author's propose an RL based mobility model for drones. The proposed model learns the fragmented 3D-coverage described in \cite{Drone2}, while trading off throughput, ping-pong, and  RLFs.   In \cite{mmWaveMobility}, authors address reliability and latency in terrestrial millimeter-wave (mmWave) mobile systems based on channel blockage prediction. In \cite{igorHandover}, authors propose reinforcement learning based approach for HO optimization. Here, authors propose threshold selection for handover parameters such as $\Delta$ and $\Gamma$ as ``action'', with reward  configuration derived from the throughput values aggregated over some duration.

\subsection{Contribution}
While the previous works for HO optimization using machine learning have focused on specific use-cases such as drones, railways, etc. or considers specific channel aspects such as mmWave, blocking, etc., there is no work to the best of our knowledge which considers the handover mobility optimization  exploiting the deployment aspects of 5G. In typical 5G stand-alone (SA) deployments, the control and synchronization are carried on much wider (large beam-width) beams called access-beams, while data for the connected UE is carried on a much narrow beams (due to beamforming) called link-beams. In typical 5G systems, the coverage for the access and link-beams are different. The link-beams are typically on mmWave with much narrow beam-width and sometimes penetrate deep into the neighbor cells, while the access-beams are in mid-band with wide beam-widths. In all the state-of-the-art HO algorithms discussed above, the HO decisions are based on the measurements done on the access-beams. In a dense 5G deployments, the connected state UE can receive access-beams from multiple BSs with sufficient power to perform initial entry procedure. In contrast to prior works, we propose a methodology to formulate HO procedure as a sub-class of reinforcement learning problem called contextual multi-arm bandit (CMAB) problem. The CMAB agent will handover UEs  opportunistically such that their average link-beam gain (and hence the throughput) can be maximized.  In the proposed system, serving BS will collect measurement reports containing beam RSRP measurements  from UEs as before, however it will not take decision on the HO, instead will forward the UE measurement reports to a centralised CMAB agent which will choose the HO action. We demonstrate the utility of such a formulation through simulations.

\section{Reinforcement Learning}
\label{sec:rl}
\begin{figure}
  \centering
\fbox{  \includegraphics[width = 3.2 in]{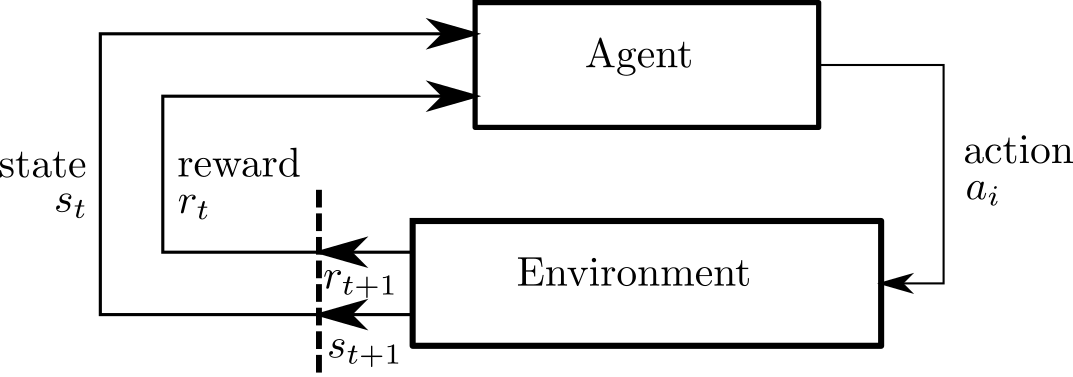}}
  \caption{Reinforcement learning framework}
  \label{fig:rl}
\end{figure}

In this Section, we will describe briefly the reinforcement learning (RL) framework. RL is an evaluative feedback based learning paradigm as shown in Fig.~\ref{fig:rl}. Here the agent learns about the optimal action in a given situation based on trial and error. This is achieved through exploration and exploitation. During exploitation the agent takes the actions that yields maximum reward, while during exploration the agent takes action which may not yield maximum reward instantaneously, however will help the agent to discover  newer actions that are profitable in the longer run.

Markov decision process (MDP) is often employed to describe RL. It is characterized by a tuple consisting of  $\left\{ \mathcal{S},\mathcal{A},\mathcal{P},\mathcal{R}\right\}$, where $\mathcal{S}$ and $\mathcal{A}$ denotes the set of possible states and actions respectively, $\mathcal{P}$ denotes the transition probabilities for the states when a particular action is taken. MDP can be solved to obtain an optimal policy, where a policy is defined as the action to be taken at each state to maximize the  reward.

\begin{figure}
  \centering
\fbox{  \includegraphics[width = 2.5 in]{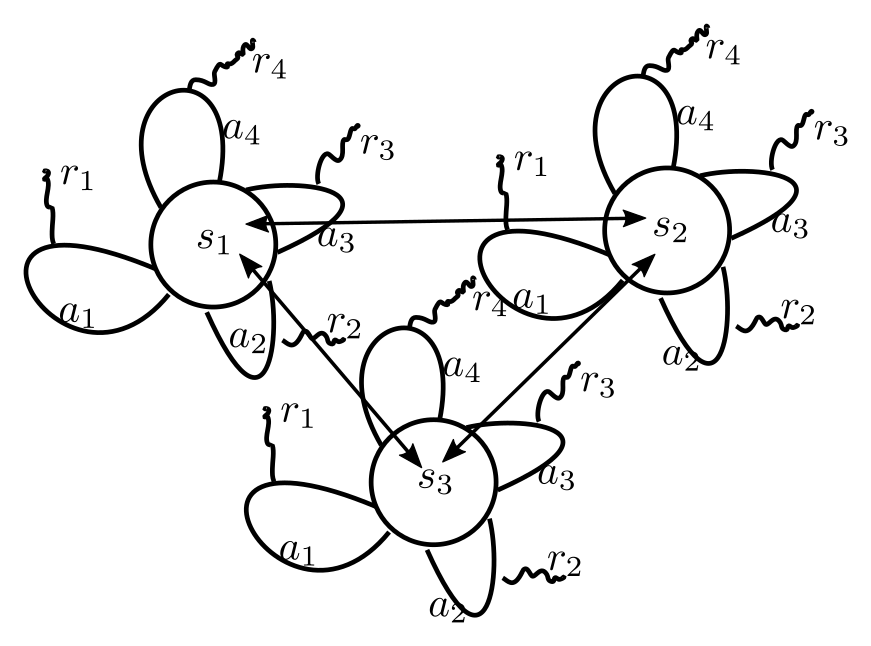}}
  \caption{CMAB as an extension of MAB. In CMAB the best arm depends on the state of the bandit.}
  \label{fig:banditMarkov}
\end{figure}

A multi-arm bandit (MAB) problem is a variant of the RL problem where the actions taken by agent does not alter the operating environment. This problem involves identifying the best arm among several arms of a muti-arm bandit whose reward distribution is unknown by trial and error. The contextual multi-arm bandit (CMAB) problem is an extension of the MAB problem in which the agent associates a context or state with the MAB. Depending on the context a particular arm yields maximum average reward. The task of the agent now is to learn the relationship between the context, arm, and reward, so that it can predict the best arm to play for a given context vector.  This is further illustrated in a $3$ context CMAB example as shown in Fig.~\ref{fig:banditMarkov}, where for each of the contexts, $s_i, i\in\{1,2,3\}$, a particular action, $a_i,i\in\{1,\ldots,4\}$, yields a better average reward.

In many practical RL problems, the model defining the $\mathcal{P}$ and $\mathcal{R}$ are not available. In these problems, an optimal policy can still be derived using Q-learning algorithm. In Q-learning method, Q-value for a policy, $\pi(a|s)$, is defined as an expected long-term reward when the agent takes an action $a$ at state $s$ and follows the policy $\pi$ thereafter. With an iterative process of exploration and exploitation, the agent can learn the optimal Q-values, $Q^*(s,a)$.  The optimal policy is to take action, $a$, which maximizes the Q-values for each state. The $\epsilon$-greedy algorithm for Q-learning is described in Algorithm~\ref{alg:cmab}. Here, the agent explores by taking random-action with probability $\epsilon$ and exploits by following an optimal-policy with probability $(1-\epsilon)$ \cite{Q-Learning}.
\begin{algorithm}[t]
  \label{alg:Qlearning}
\DontPrintSemicolon 
\caption{Q-Learning with $\epsilon$-greedy algorithm for CMAB}
\label{alg:cmab}
\KwData{$Q(s,a)\gets \mb{Initialized with random rewards}$}
\KwResult{$Q^*(s,a) \gets \mb{Optimized Q-Table}$}
$t\gets 0, s_0=s_{\mb{\scriptsize init}}$ //At time-step 0\; 
\While{$t<\mb{MAX\_STEP}$}{

  $g=\mb{rand()}$ //Random number in range [0,1] \;
  \eIf{$g<\epsilon$} {
    $a_t= \mb{rand(}\mathcal{A}\mb{)}$ //Take random action, $a_t\in\mathcal{A}$
} {
    $a_t = \underset{a}{\mb{argmax}} \{Q(s_{t},a)\}$
  }
  //Q-Table updated with average reward for $(s_t,a_t)$ combination\;
  $Q(s_t,a_t)\gets \frac{1}{T} \sum_{0}^{t}r(s_t,a_t) $// $T$ indicates number of times action, $a_t$ is taken at state, $s_t$ up to time-step $t$\;
$t \gets t+1$ \;
}
\end{algorithm}

\begin{figure}
  \centering
  \fbox{ \includegraphics[width = 2.5 in]{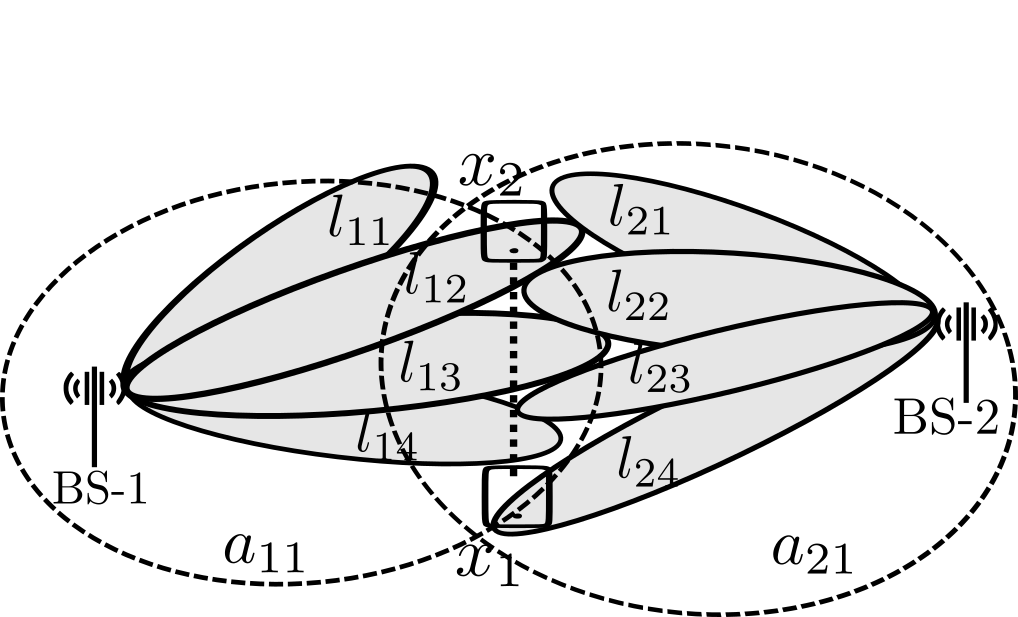}}
 \caption{An illustration of a UE moving from position $x_1$ to $x_2$ in a $2$-Node network with $1$ access-beam and $4$ link-beams.}
  \label{fig:beams}
\end{figure}


\begin{figure}[t]
\fbox{ \includegraphics[width = 3 in]{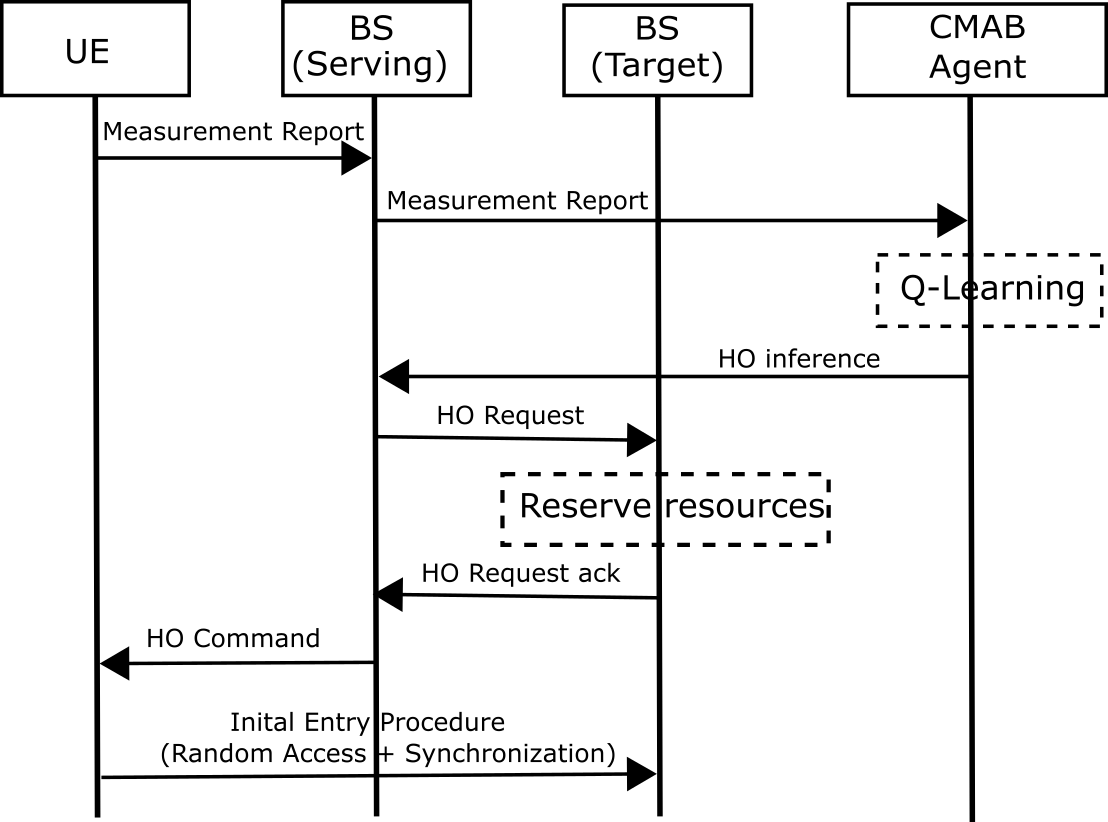}}
  \caption{HO process using centralized CMAB agent. } 
  \label{fig:HOSignaling}
\end{figure}

\section{System Aspects}
\label{sec:cmab}

In 5G, there exists two possible deployments, stand-alone (SA) and non-standalone (NSA).  In the NSA deployment, the long-term evolution (LTE) is used for cell acquisition and control, while the data is transferred using the new-radio (NR). In NR SA deployment, there are no cell specific reference signals, instead the cell acquisition is performed via synchronization signal block (SSB) beams. In this paper, we call the beams used for cell acquisition as access-beams. In 5G NR SA deployments in mmWave bands, the data is typically carried on narrow beam called link-beams.  An example illustration with $2$-BSs with $1$ access-beam and $4$ link-beams is shown in  Fig.~\ref{fig:beams}.  Note that in Fig.~\ref{fig:beams}, the access and link-beams are indicated by $a_{ij}$ and $l_{ij}$ respectively with $i$ and $j$ indicating the BS-id and beam
numbers. In the state-of-the-art HO algorithms, the wide access-beams are used in the HO inference. However, due to the dense deployment, it is possible to have access-beams corresponding to many BS are strong enough to perform HOs. Instead of using only access-beams RSRP for HO inferencing, performance can be improved by opportunistically choosing BSs for HO, which has higher link-beam gain, among  candidate BSs with sufficient access-beam power to perform initial-entry. This can be accomplished using a CMAB with link-beam RSRP after the HO action as the reward which is further explained below.

In this paper, we propose an architecture, where a centralized CMAB agent will perform HO inference. The signaling involved in the process is as shown in the  Fig.~\ref{fig:HOSignaling}. The measurements from the UEs are forwarded to the centralized CMAB agent by the serving BS.  The context for the CMAB agent consist of the access-beam measurements for serving and neighbor BSs together with the serving BS-id. Each BS can be considered as an arm. The CMAB action, i.e., pulling the arm of the bandit is analogous to choosing an appropriate BS to HO or to stay in the current BS. The goal is to select an action in a given context that maximizes the expected reward. We consider RSRP of the link-beam after HO as the reward. Since the link-beam RSRP is proportional to the throughput of the UE after HO,  the HO inference from CMAB tries to maximize the throughput.

No special measurements or signaling is needed for this method, traditional 3GPP signaling for HO as shown in Fig.~\ref{fig:HOSignaling} between BS and UE can be reused. Apart from the RSRP measurements of the access-beams for serving and neighbor cells, context for CMAB agent could also include location, speed, antenna-setup, etc.  Different reward configurations such as downlink-throughput, SINR of the link-beam, etc. after HO can also be considered.  Though we have implemented  CMAB using Q-learning in this work, it can also be implemented using algorithms such as neural network, random forest, etc.

\section{Algorithms}
\label{sec:alg}
The most common 5G HO method, is based on the RSRP measurements of access-beams. The main essence of this algorithm is that the HO is triggered by the serving BS when the access-beam RSRP of the target BS is higher than the RSRP of the serving BS  by a hysteresis value for a duration greater than time-to-trigger parameter. This algorithm runs in every BS to make an inference on whether a particular UE needs a HO. This method is briefly described in Algorithm~\ref{alg:acc}.

\begin{algorithm}[t]
\DontPrintSemicolon 
\caption{HO Algorithm using Access-Beams}
\label{alg:acc}
\small
\KwIn{Measurement Report, $\RPT$ }
\KwOut{Base station to Handover, $n$}
$\ACCNBR \gets \getNbr(\RPT)$ \;
$\ACCSER \gets \getSer(\RPT)$ \;
$\BMAX \gets \argmax(\ACCNBR)$ \;
$\TTTVAL \gets \getTTTValue()$ \;
\uIf{$\BMAX < \ACCSER + \Delta$}{
  $n \gets \getServingNode()$\;
}
\uElseIf{$\BMAX > \ACCSER + \Delta$ and $\TTTVAL < \Gamma$}{
  $\startTTTtimer()$\;
  $n \gets \getServingNode()$\;
}
\Else{
  $n \gets \getNodeId(\BMAX)$ \;
}

\Return{n}
\end{algorithm}



\begin{figure}[t]
  \centering
  \fbox{\includegraphics[width=2.8 in]{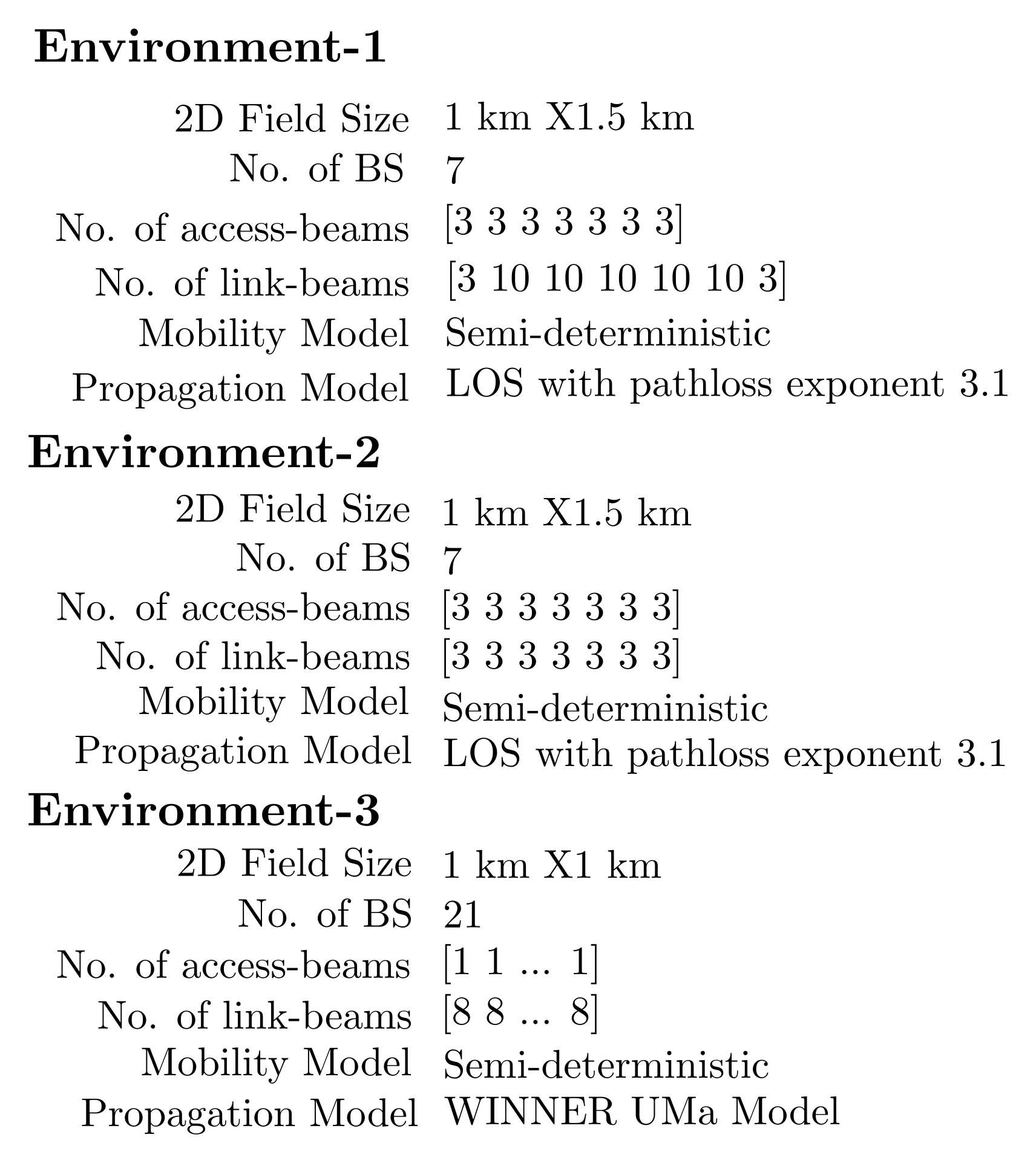}}
  \caption{The block diagram to illustrate the performance evaluation setup}
  \label{fig:env}
\end{figure}

\begin{figure*}[t!]
  \centering
  \fbox{\includegraphics[width=5.5 in]{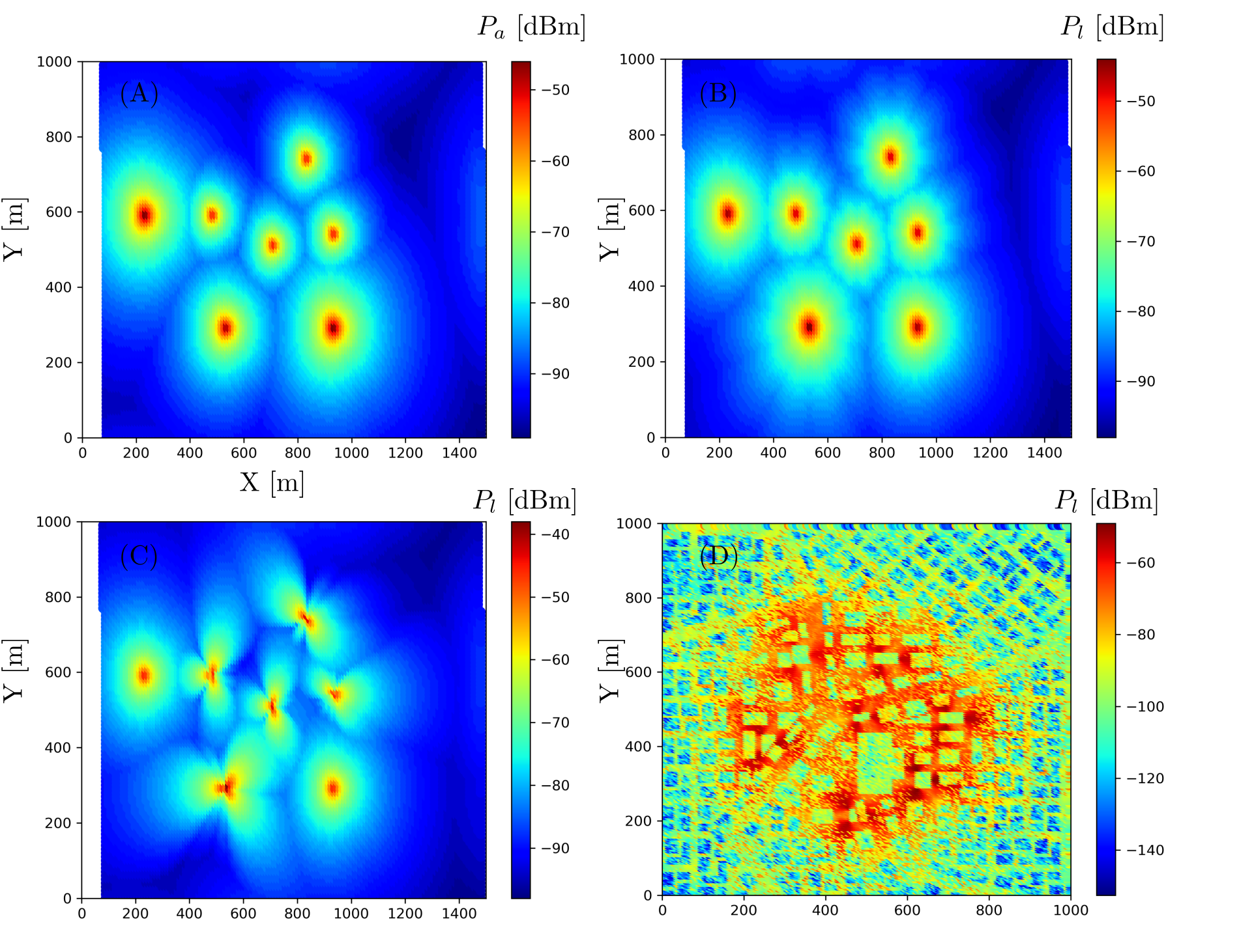}}
  \caption{The beam power density in a 2D area for different environments used in the performance evaluation. (A) shows the access-beam setup ($P_a=\max(p_{a_{ij}}), i\in[1,\ldots,7], j\in[1,2,3], \mbox{since there are 7 BS and each having 3 access-beams} $), while (B) and (C) shows link-beam energy distributions for Environment-$1$ and Environment-$2$. (D) shows the link-beam setup for the Environment-$3$}
  \label{fig:rf}
\end{figure*}


In this paper, we employ Q-learning method discussed in Algorithm \ref{alg:Qlearning}. The access-beam RSRP together with serving cell-ID  forms the ``context/state'' (refer to Fig.~\ref{fig:beams}), target BS to HO forms ``action'', and  received RSRP of the link-beam after the HO forms ``value/reward''\footnote{Value and reward are interchangeably used}.
During training phase, we set, $\epsilon=1$ in Algorithm \ref{alg:Qlearning} and the UEs are made to take random walks in a 2D radio environment. They are made to report the access-beam RSRP measurements of serving and neighbor cells which forms the context. A Q-Table is built by trying random actions for  the received states (contexts)\footnote{State and context are interchangeably used} and recording the reward (link-beam RSRP after HO) observed. During the active or online phase\footnote{Active phase is used to denote exploitation mode of the RL based HO algorithm}, we exploit the built Q-Table for optimal policy by
\begin{equation}
  \label{eq:optpolicy}
  a_t = \underset{a}{\mb{argmax}} \{Q(s_{t},a)\}.  
\end{equation}
This way of separating offline training ($\epsilon=0$) and online exploitation phase ($\epsilon=1$) will make the proposed solution practical from the 5G operational perspective by preventing the CMAB agent taking catastrophic HO action during active or live phase.

Access-beam RSRP from serving and neighbor cells which forms the context are continuous variables and  it is not possible to store all possible states/contexts in the Q-Table. Only those states which are observed during the random walk are stored in the Q-table. As long as the random walk of UEs during training phase are sufficiently long, a good representation of possible contexts are observed and corresponding state-action-values are captured in the Q-Table. During the active phase, a similarity function based on the minimum Euclidean distance measure between the Q-Table contexts and the newly received context from the measurement report can be used to choose actions from the Q-Table. This is shown in \eqref{eq:euclid}
\begin{equation}
  \label{eq:euclid}
  \mbf{c}^{\prime} = \underset{\mbf{c\in\mathcal{Q}}}{\min}\norm{\mbf{c}-\mbf{p}^{\prime}}
\end{equation}
Where $\mbf{p}^{\prime}$  denotes the received context having access-beam RSRP measurements and the serving-cell ID during the active phase. The $\mbf{c}^{\prime}$ denotes the context in the Q-Table ($\mathcal{Q}$) with minimum Euclidean distance to $\mbf{p}^{\prime}$. 

The choice of the BS during the active phase is given by
\begin{equation}
  \label{eq:val}
  i^{*}=\underset{i}{\argmax}\{V_Q \left (\mbf{c}^{\prime},C_i \right )\},
\end{equation}
where $V_Q (\mbf{c}^{\prime},C_i )$ denotes the value/reward\footnote{Value in the Q-Table is derived from the link-beam power obtained after HO} for the ``action'' of choosing the BS, $C_i$ for HO from the Q-Table for the context $\mbf{c}^{\prime}$.  The $i^{*}$ denotes the BS index with maximum reward.


Below, we illustrate the difference between the 3GPP method based on access-beam RSRP discussed in Algorithm~\ref{alg:acc}, and the  proposed CMAB method using a simple example. Consider a network with two BS in which a UE served by BS-2 is  moving from $x_1$ to $x_2$ as shown in Fig.~\ref{fig:beams}. At $x_2$, the method described in Algorithm~\ref{alg:acc} will choose to stay in BS-2, as the access-beam RSRP of BS-2 is stronger than BS-1. However the proposed CMAB agent would choose the action of HO to BS-1 because of larger reward since $l_{12}>\underset{j}{\mb{max}}(l_{2j}),j\in{1,\ldots,4}$.

\section{Results}
\label{sec:sim}
In this Section, we will evaluate the proposed method with the 3GPP access-beam based method discussed in Algorithm~\ref{alg:acc} for three distinct deployment setup. The Environment-$1$ and Environment-$2$ are based on the synthetic data generated from a system emulator with different configurations of access and link-beams. The configurations of access and link-beams for Environment-$1$ and Environment-$2$ are as shown in Fig.~\ref{fig:env}. For propagation, we used a simple path-loss model having a path-loss exponent of $3.1$. Environment-$3$ consist of $7$ roof-top sites with $21$ BSs each having $1$ access-beam and $8$ link-beams. Propagation model for Environment-$3$ is ITU standard based with  WINNER urban-macro (UMA) propagation model and is inspired by the city environment of Tokyo and Seoul \cite{winner}. The resulting RF beam patterns with the discussed configuration for the three environments are shown Fig.~\ref{fig:rf}.

As explained in the previous section, we build a Q-Table during an offline training phase by setting $\epsilon=1$ in the Q-learning method discussed in Algorithm~\ref{alg:Qlearning}. During this phase, CMAB agent takes random actions (HO to random BS) for the measurement report (context) reported by the UE while performing a random-walk in 2-D coverage area of the network. The actions which  yielded maximum reward (link-beam RSRP after HO) in a given state are retained in the Q-table after the training phase and was used in inferencing during active phase.

To access the performance in all the three environments, we employ a semi-deterministic mobility where UEs will take steps in vertical direction and when a UE hits the edge of the network, it will relocate randomly to a different X-position, and the whole process will repeat again. In each step, the measurement is sent to the CMAB agent which exploits the Q-Table using \eqref{eq:optpolicy} to make an inference on the HO decision\footnote{5G periodic measurement reporting strategy from UE to BS is employed here}. Every $10000$ steps forms an episode in our CMAB formulation. We assess the performance using the following:
\begin{itemize}
\item Average received link-beam power, $E\left(P_l\right)$ per episode
\item Probability density function (PDF) of the received link-beam power, $\mbox{p(} P_l\mbox{)}$.  
\end{itemize}

\begin{figure}[t]
  \centering
  \fbox{\includegraphics[width=3 in]{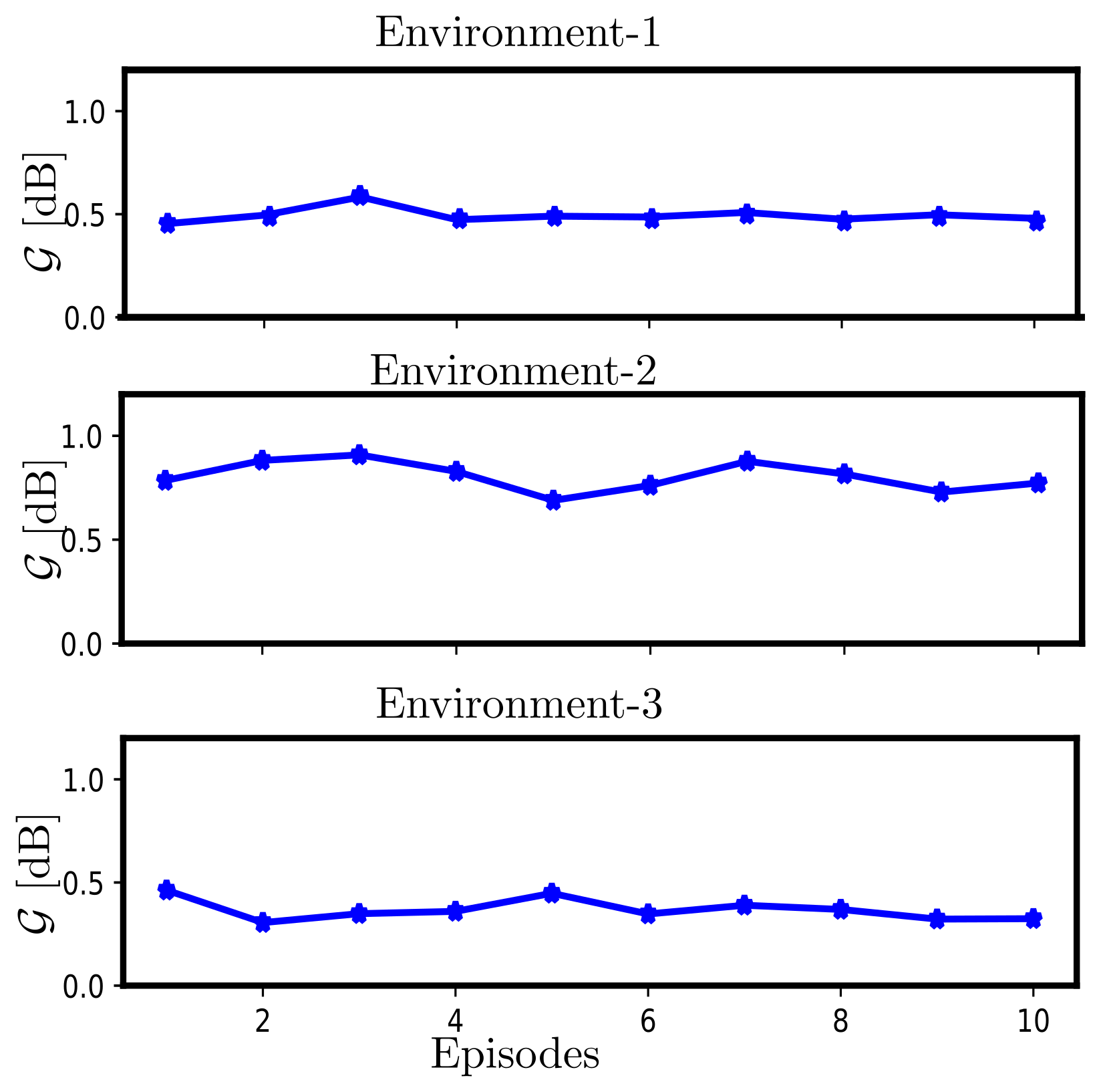}}
  \caption{The gain, $\mathcal{G}$, for 10 different episodes. Each episode constitutes $10000$ UE steps.}
  \label{fig:perfA}
\end{figure}

We compare the performance of the above metric with the access-beam based method having $\Delta$ and $\beta$ set to $0$. This will ensure a fair comparison of the 3GPP access-beam based algorithm with the proposed RL algorithm, since RL-algorithm does not penalize the ping-pong during HO. We define gain, $\mathcal{G}$, as increase in the average link-beam gain by using CMAB, and is defined as 

\begin{equation}
  \label{eq:G}
  \mathcal{G}=\{E\left(P_l\right)\}_{\mbox{\scriptsize Algorithm-}\ref{alg:Qlearning}}-\{E\left(P_l\right)\}_{\mbox{\scriptsize Algorithm-}\ref{alg:acc}},
\end{equation}
where Algorithm-\ref{alg:acc} is the 3GPP HO algorithm based on access-beam RSRP and the Algorithm-\ref{alg:Qlearning} is the proposed CMAB based HO algorithm. The gain, $\mathcal{G}$, for different episodes with different initialization points are given in Fig.~\ref{fig:perfA}.
\begin{figure}[t]
  \centering
  \fbox{\includegraphics[width=3 in]{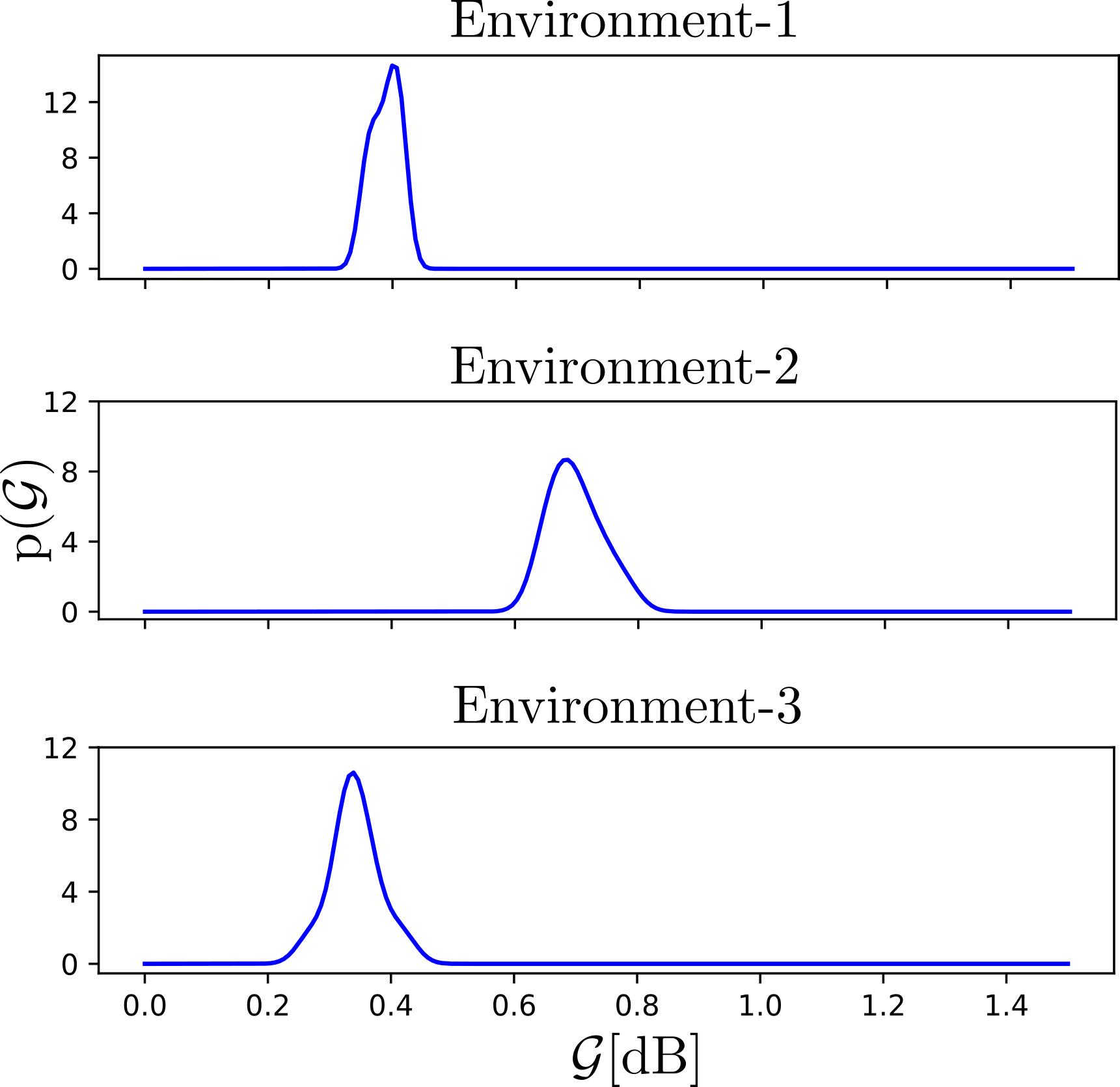}}
  \caption{The PDF of gain, $\mathcal{G}$ for different environments.}
  \label{fig:perfB}
\end{figure}
\begin{figure}[t]
  \centering
  \fbox{\includegraphics[width=3 in]{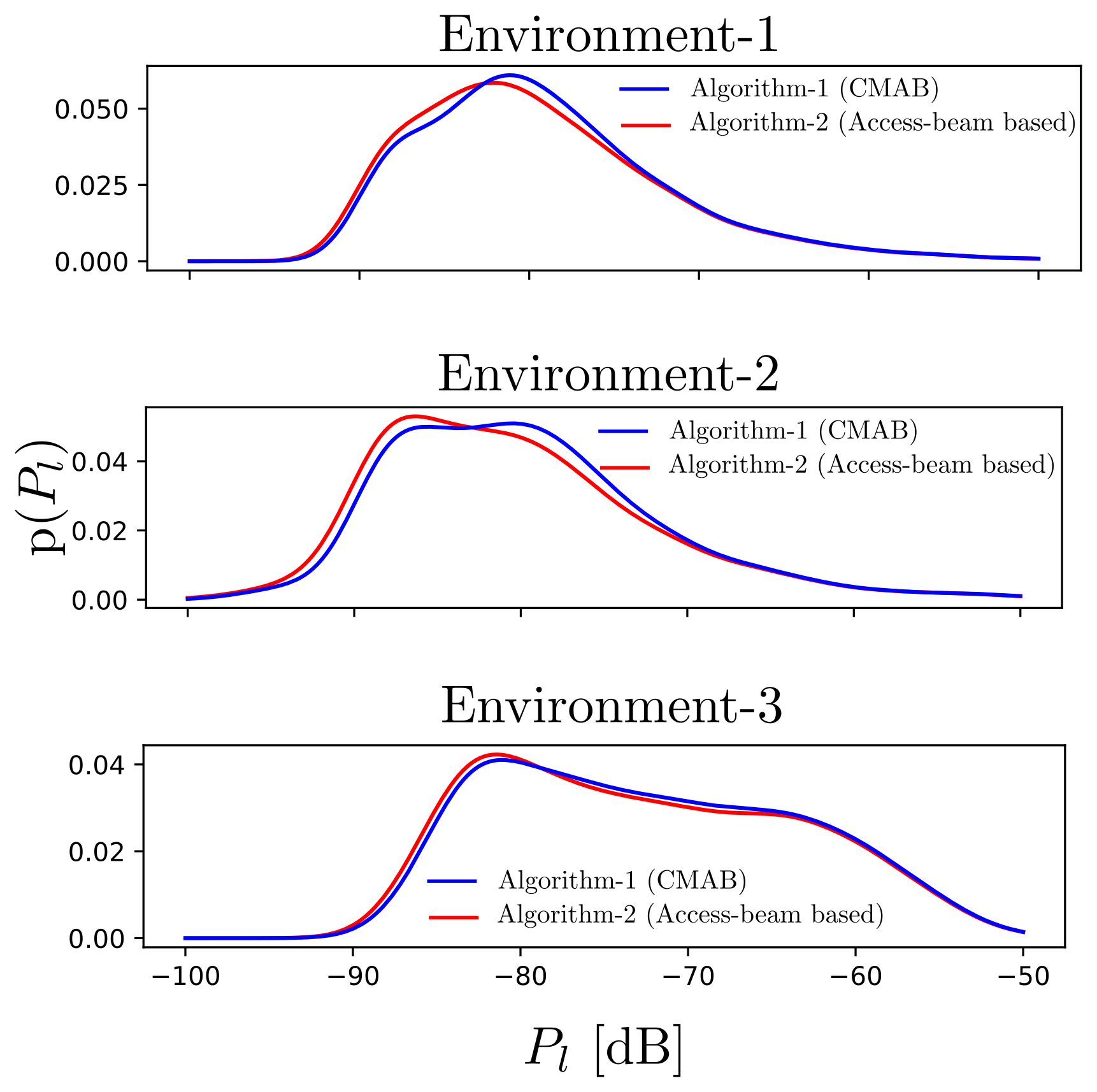}}
  \caption{The PDF of link-beam RSRP, $P_l$, for all three environment.}
  \label{fig:perfC}
\end{figure}
Notice from Fig~\ref{fig:perfA} that the gain, $\mathcal{G}$, is positive for all the episodes in all three environments. 
The PDF of the $\mathcal{G}$ is shown in Fig.~\ref{fig:perfB}. Notice that the gain, $\mathcal{G}$ in the Environment-$2$ is more than in Environment-$1$ this is due to the higher opportunity for RL based handover to pick better BSs as the link-beams in this environment are narrow and penetrate deep into the neighbor cells (refer to Fig.~\ref{fig:rf}). For Environment-3 which is based on the WINNER UMA propagation model, the results indicate a gain between $[0.3 \mbox{ - }0.5]~\mb{dB}$. The PDF of the link-beam RSRP, $P_l$, experienced by the UE using both the algorithms are shown in Fig.~\ref{fig:perfC}. Notice that the distribution of $p(P_l)$ for CMAB algorithm is shifted to the right for the proposed method indicating the improvement in the link-beam performance.  Since the link-beams carry  physical downlink shared channel (PDSCH) data, the improvement in link-beam RSRP will increase the downlink throughput for the UE. The quantum of improvement depends on among other things,  channel condition, interference perceived by the UE.

\section{Conclusion and Discussion}
\label{sec:conclusion}
In this paper, we proposed a HO algorithm for 5G system using RL. We showed that the HO problem can be posed as a sub-class of RL problems called CMAB.  We showed how such a system can be developed using a Q-learning method. We also discussed mitigation strategies for some of the challenges of the design such as state-space explosion by building a Q-Table with representative states during the training phase and with suitable choice of similarity function to pick the closest state in Euclidean space during the active phase for HO inference.

We assessed the performance for different deployment and propagation environments including an ITU standard based one. We demonstrated the utility of the method through average link-beam performance using a semi-deterministic mobility model in three distinct environments. In all the considered environments, the proposed method of this paper performs better than the existing methods. The results also indicate that when the link-beams are narrow and penetrate deep into the neighbor cells, which will be common in dense 5G cellular deployments in mmWave band, the RL based HO algorithm performs better due to the increased opportunity to optimize long-term link gains.

\section*{Acknowledgements}
We would like to thank Ankit Jauhari and Mirsad Cirkic of Ericsson Research for sharing their insights. The discussions with them on  Q-Learning and 5G deployments were helpful in improving the paper. 

\bibliography{main}
\bibliographystyle{IEEE}
 
\end{document}